\documentclass[11pt]{article}
\usepackage{amssymb,amsmath}
\usepackage{mathtools}
\usepackage{cancel}

\usepackage{tikz-cd}
\usepackage{xcolor}
\usepackage[normalem]{ulem}

\newdimen\tableauside\tableauside=1.0ex
\newdimen\tableaurule\tableaurule=0.4pt
\newdimen\tableaustep
\def\phantomhrule#1{\hbox{\vbox to0pt{\hrule height\tableaurule
width#1\vss}}}
\def\phantomvrule#1{\vbox{\hbox to0pt{\vrule width\tableaurule
height#1\hss}}}
\def\sqr{\vbox{%
  \phantomhrule\tableaustep

\hbox{\phantomvrule\tableaustep\kern\tableaustep\phantomvrule\tableaustep}%
  \hbox{\vbox{\phantomhrule\tableauside}\kern-\tableaurule}}}
\def\squares#1{\hbox{\count0=#1\noindent\loop\sqr
  \advance\count0 by-1 \ifnum\count0>0\repeat}}
\def\tableau#1{\vcenter{\offinterlineskip
  \tableaustep=\tableauside\advance\tableaustep by-\tableaurule
  \kern\normallineskip\hbox
    {\kern\normallineskip\vbox
      {\gettableau#1 0 }%
     \kern\normallineskip\kern\tableaurule}%
  \kern\normallineskip\kern\tableaurule}}
\def\gettableau#1 {\ifnum#1=0\let\next=\null\else
  \squares{#1}\let\next=\gettableau\fi\next}

\def\be{\begin{equation}}
\def\ee{\end{equation}}
\def\ba{\begin{array}}
\def\ea{\end{array}}

\def\dd{\partial}

\newenvironment{al}{\subequations\align}{\endalign\endsubequations}
\newenvironment{all}{\equation\aligned}{\endaligned\endequation}

\newcommand{\ol}[1]{\overline{#1}}

\def\oD{\ol\nabla}
\def\0ads{AdS$_d$}
\def\1ads{AdS$_{d+1}$}

\def\3ads{AdS$_3$}
\def\2ads{AdS$_2$}

\def\Fa{Fronsdal action }
\def\Ff{Fronsdal field }

\def\gt{the gauge transformation }

\newcommand{\bref}[1]{\textbf{\ref{#1}}}

\usepackage{accents}
\usepackage{graphicx}

\DeclareMathOperator{\sech}{sech}

\makeatletter
\newcommand{\douwidehat}[2]{%
  \sbox0{$\m@th#1\widehat{\hphantom{#2}}$}%
  \sbox2{$\m@th#1x$}
  \sbox4{$\m@th#1#2$}
  \dimen0=\ht0
  \advance\dimen0 -.8\ht2
  \dimen2=\dp4
  \rlap{%
    \raisebox{\dimexpr\dimen0-\dimen2}{%
      \scalebox{1}[-1]{\box0}%
    }%
  }%
  {#2}%
}
\usepackage{jheppub}

\title{AdS$_3$/AdS$_2$ degression of Fronsdal fields}

\author[a]{Alexander\ Yan}

\affiliation[a]{I.E. Tamm Department of Theoretical Physics, \\P.N. Lebedev Physical
Institute,\\ Leninsky ave. 53, 119991 Moscow, Russia}
\emailAdd{yan@lpi.ru}

\abstract{We analyze the Kaluza--Klein type procedure in AdS$_3$ space called the dimensional degression. The topological theory of the Fronsdal field in AdS$_3$ is reformulated in terms of the fields propagating in AdS$_2$. We find that the Fronsdal field in AdS$_3$ leads to finitely many Kaluza--Klein modes. Namely, the obtained spectrum is the massive Klein--Gordon and Proca fields in AdS$_2$. The result is derived by using  the specific mode expansion, gauge fixing, and 2--dimensional Schouten identities.}

\begin{document}

\allowdisplaybreaks 

\maketitle
\flushbottom

\section{Introduction}\label{s:intro}

We study the degrees of freedom encoded in the Fronsdal theory in \3ads space \cite{Fronsdal:1978rb}. This theory  can be equivalently described as $3d$ Chern--Simons theory with $sl(N)\oplus sl(N)$ algebra \cite{Blencowe:1988gj, Campoleoni:2010zq, Campoleoni:2011tn}, that explains  why the theory is topological. Nevertheless, being a topological theory in $3$ dimensions does not necessarily mean that there are no degrees of freedom at all. In fact, it indicates that  solutions of the equations of motion (EOM) (in our case second order PDE) are not functions on a $2$--dimensional hypersurface as prescribed by the Cauchy--Kovalevskaya theorem and as we observe in non--topological theories. Indeed, in topological theories  solutions can have functional freedom but on a hypersurface with lesser than $2$ dimensions. For instance, the general solution of the 3D Einstein equations with a negative cosmological constant found by Banados contains two arbitrary functions on the 1--dimensional hyperspace \cite{Banados:1998gg}. The degrees of freedom of the \3ads Fronsdal theory  can be seen from the \2ads perspective. To do that, we consider the AdS$_3$/AdS$_2$ dimensional degression \cite{Metsaev:2000qb, Artsukevich:2008vy, Gwak:2016sma, Alkalaev:2021zda}. The degression allows us to yield the EOM in \2ads whose solutions are functions on the 1--dimensional hypersurface. Thus, the degrees of freedom of the Fronsdal field can be revealed.

For the sake of generality and convenience of calculations, let us discuss the AdS$_{d+1}$/AdS$_d$ dimensional degression for arbitrary $d$. In general, the AdS$_{d+1}$/AdS$_d$ dimensional degression is the Kaluza--Klein procedure with the following metric \cite{Gutperle:2020gez}
\begin{equation}\label{metric}
    ds(\text{AdS}_{d+1})^2  =  \cosh^2z \; ds(\text{AdS}_{d})^2  + dz^2\,.
\end{equation}
The metric represents the \1ads space sliced into the \0ads spaces. There is no compact direction in \1ads so the common term ``compactification'' is not suitable here. However, one can perform the Kaluza--Klein mode expansion along the slicing direction $z$, integrate out the $z$--dependence and obtain the theory in \0ads. Thus, the degression reformulates the \1ads theory as the theory in \0ads. The degrees of freedom encoded in the theory are transferred from the \1ads to \0ads background.

In higher dimensions $d+1\geqslant 4$ the \1ads/\0ads  degression provides infinitely many Kaluza--Klein modes. Roughly speaking, an arbitrary function on the $d$--dimensional hypersurface in \1ads is identically reformulated as an infinite Taylor series of arbitrary functions on $(d-1)$--dimensional hypersurface in \0ads. Infinitely many degrees of freedom in \0ads can be interpreted as a finite number of degrees of freedom in \1ads that characterizes non--topological theory. A radically different picture is observed when the $d+1=3$ case is considered. In this paper we show that the Fronsdal theory in AdS$_3$ is described with finitely many Kaluza--Klein modes. Finitely many degrees of freedom in \2ads can not be interpreted as a degree of freedom in \3ads. Thus, the Fronsdal theory in AdS$_3$ is indeed topological.

This paper continues the analysis started in our previous work \cite{Alkalaev:2021zda}, where we showed that the totally--symmetric massless irreducible representation $\mathcal{H}(s,s)$ (with integer spin $s\geqslant 2$)  of ${o(2,2)}$ algebra branches into two massive irreducible representations $\mathcal{D}_s\oplus \mathcal{D}_s$ of ${o(2,1)}$ algebra. The same result was obtained from the EOM perspective on $s=2,3$ examples. The present paper considers the $s\geqslant 4$ case. Analyzing the action and EOM, we proved that only two Kaluza--Klein modes propagate in \2ads. They constitute two massive irreducible representations
\begin{all}
&\text{AdS$_2$ Klein-Gordon field:}\qquad\;\;&&\Big[\Box - s(s - 1)\Big]\varphi = 0\;,\\
&\text{AdS$_2$ Proca field:} &&\Big[ \Box -\big(s(s-1)-1\big)\big]A^{\mu} = 0\;, &&\nabla_\mu A^\mu = 0\;,
\end{all}
where the masses (or, equivalently, the energies $E=s$\footnote{The mass and energy values are related in \0ads as $m^2 = E(E-d+1)-n$, where $n$ is the spin of the irreducible representation, see e.g. \cite{Metsaev:1997nj}.}) indeed coincide with prescribed values obtained in the group--theoretical approach.

Let us outline the basic concepts that will be addressed  in the paper. Note that the dimension $d$ is kept  arbitrary until the last step when the EOM is analyzed. The Fronsdal theory describes a  massless spin-$s$ bosonic field $\Phi^{m(s)}(x,z)\equiv \Phi^{(m_1 ...m_s)}$ propagating in the AdS$_{d+1}$ background with the following action
\begin{align}\label{e:Fa}
\begin{split}
      S = &\int d\mu_{d+1}\Big\{ -(\oD_m\Phi^{n(s)})^2 + s(\oD_m\Phi^{mn(s - 1)})^2 - s(s - 1)\oD_m\Phi'_{n(s - 2)}\oD_k\Phi^{mn(s - 2)k} + \\
  &+\frac{s(s -1)}{2}(\oD_m\Phi'^{n(s - 2)})^2 + \frac{s(s - 1)(s - 2)}{4}(\oD_m \Phi'^{mn(s - 3)})^2 - \\
  & -\big((s - 2)(d + s - 2) - s\big)(\Phi^{m(s)})^2 + \frac{s(s - 1)\big((s - 1)(d + s - 2) - 2\big)}{2}(\Phi'^{m(s-2)})^2  \Big\} \;,
\end{split}
\end{align}
where a prime denotes the trace with respect to the \1ads metric $\Phi'^{m(s-2)} = {\Phi^{m(s-2)nk}}\ol g_{nk}$ and $\ol g_{nk}(x,z)$ is the \1ads metric \eqref{metric}. The Fronsdal field is subjected to the double--tracelessness condition $\Phi''^{\mu(n-4)} = 0$. See Appendix \bref{a:Con_not} for the list of our conventions. The Fronsdal action is invariant under the following gauge transformations
\begin{equation}\label{e:1gauge}
  \delta\Phi^{m(s)} = \oD^{m}\Xi^{m(s - 1)}\;,
\end{equation}
where the gauge parameter $\Xi^{m(s-1)}$ is a traceless totally--symmetric tensor.

Our analysis faces two main difficulties. The first one relates to an arbitrary number $s$ of world indices. Making the $d+1$ split, the Fronsdal field $\Phi^{m(s)}$ in \1ads decomposes into $s+1$ fields $\phi^{\mu(n)}$ (the index $\mu$ is associated with \0ads). The straightforward approach is to perform the $d+1$ split in the action. However, this way is technically difficult. Relying on our  experience in the analysis of the $s=2,3$ cases, we can anticipate the final result and construct an ansatz action with arbitrary constant coefficients which will be fixed by the gauge invariance. The second difficulty is related to the gauge fixing. Most of the fields $\phi^{\mu(n)}$ in the \0ads theory are not dynamical. In order to find the spectrum of the theory, the fields have to be partially gauge fixed. A convenient choice of a gauge -- the traceless--transverse (TT) gauge -- turns out to be  inconsistent with the EOM if imposed directly. So, the consistency condition of the TT  gauge has to be verified on--shell.

The organization of the paper repeats the general stages of the degression procedure. In Section \bref{s:def} we perform the $d+1$ split of the \1ads Fronsdal field and then construct component fields $\phi^{\mu(n)}$ and gauge parameters $\xi^{\mu(n)}$ meeting the (double)--tracelessness requirement. Note that e.g. the spin--2 field $\phi^{\mu(2)}$ does not obey the double--tracelessness  and hence can be chosen differently. The arbitrariness of $d$ allows us to make calculations shorter by selecting the field $\phi^{\mu(2)}$ in a form analogous to higher--spin counterparts $\phi^{\mu(n)}$. On the one hand, with this choice the ansatz component actions can be constructed in a unified form for all field sectors. On the other hand, this choice creates a pole $\frac{1}{d-2}$ at $d=2$ in the spin--0,2 sectors. So, after computing the gauge transformation in terms of the component fields and gauge parameters and finding the constant coefficients of the ansatz we redefine the spin--2 field to remove the pole. Finally, we find the component actions and gauge transformations without the pole provided that the redefinition is done. In Section \bref{s:elim} we introduce the specific Kaluza--Klein mode expansion to integrate out the $z$--coordinate. The mode expansion uses basis functions built of the Jacobi polynomials. The $z$--dependence is eliminated from most of the component actions by using the properties of the Jacobi polynomials. All but two spin--0,1 field modes are the Stueckelberg fields. Gauge fixing them,  we integrate out the slicing coordinate $z$ in the rest of the component actions. In this way, we obtain the \0ads theory. In Section \bref{s:EOM} we analyze the EOM and, finally, fix $d=2$. In 2--dimensional spaces there are the so--called Schouten identities which allow one to substitute kinetic terms in the EOM with algebraic ones like  $\Box\phi^{\mu(n)}\sim \phi^{\mu(n)}$. It turns out that the EOM decomposes into three separate non--interacting field sectors:  spin--0 and spin--2, spin--1, and higher--spin sectors. Only the first two sectors are dynamical. The latter one requires additional treatment to impose the TT gauge. In Appendix \bref{a:Con_not} we collect our conventions and notation. In Appendix \bref{s:Jac} we consider the basis functions built of the Jacobi polynomials. In Appendix \bref{s:how} we discuss  the ansatz action in more detail. In Appendix \bref{ap:An-In} we list explicit expressions for various cumbersome coefficients in the ansatz action. In Appendix \bref{s:s4} we analyze the higher--spin sector of the EOM in the case of the spin $s=4$ in order to illustrate calculations in Section \bref{s:EOM}. 

\section{Component actions and gauge transformations}\label{s:def}

In this Section we reformulate the \1ads Fronsdal action \eqref{e:Fa} and gauge transformation \eqref{e:1gauge} in terms of objects in \0ads parametrized by the slicing coordinate $z$.  

\subsection{Defining component fields and gauge parameters}

We start with defining component fields $\phi^{\mu(n)}$ and gauge parameters $\xi^{\mu(n)}$ constructed from $\Phi^{m(s)}$ and    $\Xi^{m(s-1)}$, respectively. Since we want to interpret $\phi^{\mu(n)}$ as Fronsdal fields in \0ads, they are chosen to be totally--symmetric and double--traceless with respect to the \0ads metric. The same logic is applied to the component gauge parameters $\xi^{\mu(n)}$ which are to be totally--symmetric and traceless with respect to the \0ads metric as well. 

 Firstly, we perform the $d+1$ split. Namely, indices of the \Ff $\Phi^{m(s)}$ are split into the Greek ones and the slicing one $m=(\mu,\bullet)$. Thus, we obtain $s+1$ tensors $\Phi^{\mu(n)\bullet(s-n)}$ which are not double--traceless with respect to the \0ads metric by itself (our symmetrization rules are given in \eqref{sym_rule}). To achieve the double-tracelessness one has to add terms $g^{\mu(2)}\Phi^{\mu(n-2)\bullet(s-n+2)}$, $g^{\mu(2)}g^{\mu(2)}\Phi^{\mu(n-4)\bullet(s-n+4)}$, etc. So, the component Fronsdal fields are 
 \begin{equation}\label{e:comp_fiel}
  \phi^{\mu(n)} = \sum^{\left[n/2\right]}_{k=0}\alpha^k_n (\sech z)^{2k} g^{\mu(2)}\ldots g^{\mu(2)} \Phi^{\mu(n-2k)\bullet(s-n+2k)}\;,
\end{equation}
where
\begin{equation}\label{e:alpha}
  \alpha^k_n = \frac{\Gamma(\frac{d}{2}+n-2-k)}{2^k\Gamma(\frac{d}{2}+n-2)}\;,
\end{equation}
for $0 \leqslant n \leqslant s$ and $0\leqslant k\leqslant [\frac{n}{2}]$. The coefficients $\alpha^k_n$ are defined uniquely up to the overall factor for $4\leqslant n\leqslant s$ from the double--tracelessness. In all the other cases $0 \leqslant n \leqslant 3$ there is no double--tracelessness condition and, in general, spin--$0,1,2,3$ fields could be defined in a different way. But we chose the definition \eqref{e:comp_fiel}, \eqref{e:alpha} to have a unified form for it leads to much simpler computations of the gauge transformations and action in terms of the component fields and gauge parameters.   

The component gauge parameters are treated in the same fashion. The coefficients in linear combinations are chosen in such a way that the gauge parameters $\xi^{\mu(n)}$ are to be traceless
\begin{all}\label{def_xi}
  &\xi^{\mu(n)} = \sum^{\left[n/2\right]}_{k=0}\beta^k_n (\sech )^{2k-2}g^{\mu(2)}\dots g^{\mu(2)} \Xi^{\mu(n-2k)\bullet(s-1-n+2k)}\;.
\end{all}
for $0 \leqslant n \leqslant s-1$ and $0\leqslant k\leqslant [\frac{n}{2}]$. It turns out that the coefficients here have a relation with the previous ones \eqref{e:alpha} $\beta^k_n = \alpha^k_{n+1}$. 

The price to pay for convenient computations is that the definition \eqref{e:comp_fiel}, \eqref{e:alpha} contains a pole at $d=2$ in the spin--2 field
\begin{equation}\label{e:pole}
    \alpha^1_2  = \frac{1}{d-2}\;.
\end{equation}
This issue is not novel and had already been discussed in \cite{Alkalaev:2021zda}. Note that the other fields and gauge parameters do not provide the analogous problem and so do not require additional treatment. To circumvent the issue, we perform calculations in Section \bref{s:ansatz} assuming $d$ to be arbitrary. Then, in Section \bref{s:redef}, the pole will be removed by a spin--2 field redefinition
\begin{equation}\label{redef}
    \widetilde\phi^{\mu(2)} = \Phi^{\mu(2)\bullet(s-2)}\;.
\end{equation}



\subsection{Gauge transformation and ansatz action}\label{s:ansatz}

Using the definitions \eqref{e:comp_fiel}, \eqref{def_xi}, one can rewrite the gauge transformation \eqref{e:1gauge} in terms of the component fields and gauge parameters 
\begin{equation}\label{e:gauge_L}
  \delta \phi^{\mu(n)} = \nabla^{\mu}\xi^{\mu(n-1)} + (s-n)L_{\downarrow}\xi^{\mu(n)} + \frac{2(d+n+s-4)}{(d+2n-6)(d+2n-4)}L_{d+2n-4}g^{\mu(2)}\xi^{\mu(n-2)}\;,
\end{equation}
where $L_{\downarrow}$ and $L_{d+2n-4}$ are the lowering/raising operators\footnote{We also call them the $L$--operators. The term ``lowering/raising'' in the context of the factorization of a second--order differential operator into a product of two first--order differential operators was proposed in \cite{RevModPhys.23.21}. The role of these operators for constructing the ansatz action is discussed in Appendix \bref{s:how}.}, respectively, defined in \eqref{def_L}. By default, here and in the sequel when a term cannot exist by a simple count of indices, we imply that it is absent.\footnote{ Namely, in \eqref{e:gauge_L} the first, the second, and the last terms are absent for $n=0$, $n=s$, $n=0,1$, respectively.} A few comments are in order. 
\begin{enumerate}
    \item  Since the last term exists only for $n\geqslant 2$, the gauge transformations do not possess poles other than \eqref{e:pole}.
    
    \item By looking at the original \1ads gauge transformation \eqref{e:1gauge}, the tensorial structure of \eqref{e:gauge_L} can be anticipated. The three terms are the only terms that can be written with one derivative at most, double--tracelessness condition and without index contraction. 
    
    \item The first term $ \delta \phi^{\mu(n)} = \nabla^{\mu}\xi^{\mu(n-1)}$ describes the Fronsdal gauge transformation part. That is why we call $\phi^{\mu(n)}$ the Fronsdal fields. The other two terms point to the interacting behavior of a system of the Fronsdal fields $\phi^{\mu(n)}$. Equation \eqref{e:gauge_L} implies that in order to be gauge invariant the spin--$n$ field EOM can contain only component fields of spin from $n-2$ to $n+2$ at most with an appropriate number of the $L$--operators. This idea will be used to construct the ansatz action, see Appendix \bref{s:how}.   
\end{enumerate}

The detailed analysis of the spin--$2,3$ case in the previous work \cite{Alkalaev:2021zda} gave us enough experience to construct an ansatz action with arbitrary coefficients. This ansatz action is not the most general one and contains only relevant terms which do not vanish after verifying the gauge invariance. Such an educated guess results from the $d+1$ split and simplifies cumbersome calculations. 

The ansatz action has the following form
\begin{equation}\label{sumS}
    S = \sum^s_{n=0}\Big( S_{nn}+S_{nn-1}+S_{nn-2}\Big)\;,
\end{equation}
where $S_{ij}$ are called the component actions. Two indices of a component action represent ranks of two fields contracted within this action.\footnote{Only those component actions exist where $n-1$ or $n-2$ are non--negative.\label{foot:ansatz}} The component actions read
\begin{al}
  S_{nn} =  &\iint (\cosh z)^{2(n-1)} \Big\{ A_n\Big[ -(\nabla_\mu \phi^{\nu(n)})^2 + n(\nabla_\mu\phi^{\mu\nu(n-1)})^2 -  n(n-1)\nabla_\mu \phi'_{\nu(n-2)}\nabla_\rho \phi^{\mu\nu(n-2)\rho} + \notag\\
  & +\frac{n(n-1)}{2}(\nabla_\mu \phi'^{\nu(n-2)})^2 + \frac{n(n-1)(n-2)}{4}(\nabla_\mu\phi'^{\mu(n-3)})^2   \Big] +  B_n(\phi^{\mu(n)})^2 + C_n(\phi'^{\mu(n-2)})^2  + \notag\\
  & + D_n\phi_{\mu(n)}L_{d+2n-4}L_{\downarrow}\phi^{\mu(n)} + E_n\phi'_{\mu(n-2)}L_{d+2n-4}L_{\downarrow}\phi'^{\mu(n-2)}\Big\}\;,\label{e:ansatz_nn}\\
  S_{nn-1} = &\iint (\cosh z)^{2(n-1)}\Big\{ F_n\nabla_\mu\phi^{\mu\nu(n-1)}L_{d+2n-4}\phi_{\nu(n-1)} + G_n\nabla_\mu\phi'_{\nu(n-2)}L_{d+2n-4}\phi^{\mu\nu(n-2)} +\notag\\
  &+ H_n\nabla_\mu\phi'^{\mu\nu(n-3)}L_{d+2n-4}\phi'_{\nu(n-3)}\Big\}\;,\\
  S_{nn-2} = &\iint (\cosh z)^{2(n-1)} I_n\phi'^{\mu(n-2)}L_{d+2n-4}L_{d+2n-6}\phi_{\mu(n-2)}\;,\label{e:ansatz_nn-2}
\end{al}
where the double integral notation is introduced in \eqref{double_int}. From now on the prime represents the trace with respect to the \0ads metric $\phi'^{\mu(n-2)} = \phi^{\mu(n-2)\nu\rho}g_{\nu\rho}$. 

The requirement of the gauge invariance fixes the coefficients $A_n,\ldots,I_n$ for all $0\leqslant n \leqslant s$.\footnote{In fact, the gauge invariance does not fix the overall factor. But it can be proved from the general consideration of the $d+1$ split of the action that the first coefficient in the highest sector is $A_s=1$. } The expressions for the coefficients are listed in Appendix \bref{ap:An-In}.  A few arguments on how to guess this form of the ansatz action are given in Appendix \bref{s:how}. Note that constants $A_0$, $B_0$, $F_1$, $I_2$ in the lower sector contain the pole \eqref{e:pole}. It is the only pole in the action. Intermediate steps do not produce any new poles.

\subsection{Redefinition of spin--2 field}\label{s:redef}

To get rid of the pole \eqref{e:pole} at $d=2$ we redefine the spin-2 component field
\begin{equation}\label{e:redef}
    \widetilde\phi^{\mu(2)} \equiv \phi^{\mu(2)} - \frac{1}{d-2}\sech^2 z\;g^{\mu(2)}\phi\;.
\end{equation}
This redefinition is exactly the same as if we had identified the spin--2 component field with \eqref{redef} from the beginning. Having made the definition \eqref{redef}, the pole \eqref{e:pole} would not appear in all following calculations. Note that these two different definitions \eqref{e:comp_fiel} and \eqref{redef} of the spin--2 component field is just an identical mixing of rank--0 terms $\Phi^{\bullet(s)}$ in the action \eqref{e:Fa} after the $d+1$ split. Thus, to obtain the resulting action with the definition \eqref{redef} or to calculate the action with the definition \eqref{e:comp_fiel} and then make the redefinition \eqref{e:redef} are two equivalent ways. 

The spin-2 field $\phi^{\mu(2)}$ is contained only in $S_{42}$, $S_{32}$, $S_{22}$, $S_{21}$, $S_{20}$ component actions. After the redefinition each component action with the spin--2 field produces modified spin--2 terms plus new spin--0 terms, e.g. $S_{42} = \tilde S_{42} +  S_{40}$. Thus, the redefinition  affects the spin--0 sector. So, the following component actions $S_{20}$, $S_{10}$, $S_{00}$ are changed and two new terms $S_{40}$, $S_{30}$ are produced\footnote{Those component actions that changed only by acquiring the modified spin--2 field ($\tilde S_{42}$, $\tilde S_{32}$, $\tilde S_{22}$, $\tilde S_{21}$) are not listed here. In the sequel, the tilde will be omitted.}
\begin{al}
    S_{40} = & 0\;, \qquad\qquad\qquad\qquad  S_{30} = \iint\cosh^2z\;U_1 \nabla_\mu \phi'^{\mu}L_d\phi\;,\label{e:act2_a}\\
  S_{20} =& \iint \Big\{ W_1 \big[ \nabla_\mu\nabla_\nu \phi^{\mu\nu} - \Box\phi' \big] \phi + \cosh^2z\;\phi'\Big[  X_1 \dd^2 +X_2\tanh z\dd + X_3 \tanh^2z + X_4 \Big]  \phi \Big\} \;,\label{e:act2_b}\\
  S_{10} =& \iint  \nabla_\mu \phi^{\mu}\big( V_1 \dd + V_2 \tanh z \big)\phi\;,\label{e:act2_c}\\
  S_{00} =& \iint\frac{1}{\cosh^2z}\Big\{ Y_1 (\nabla_\mu \phi)^2 + \phi\Big[ Z_1 \dd^2 +Z_2 \tanh^2z + Z_3  \big)  \Big]\phi \Big\}\;,\label{e:act2_d}
\end{al}
where $\dd = \frac{\dd}{\dd z}$ is the $z$--derivative, see Appendix \bref{s:Jac}.\footnote{There is no such term as $\iint \phi \tanh z\dd \phi$ in $S_{00}$ by virtue of the relation
\begin{equation}
  \int dz \cosh^dz A(z)B(z)\dd A(z) = -\frac{1}{2}\int dz \cosh^dz\; A^2(z)L_dB(z)
\end{equation}
which follows from integration by parts.} The constants $U_1,\ldots,Z_3$ are determined in Appendix \bref{ap:An-In}. The component action $S_{40}$ vanishes since it contains the double trace of the spin--4 field which is double--traceless.

The field redefinition \eqref{e:redef} affects not only the component actions but also the spin--2 gauge transformation \eqref{e:gauge_L}
\begin{equation}\label{e:gauge_L2}
  \delta \phi^{\mu(2)} = \nabla^{\mu}\xi^{\mu} + (s-2)L_{\downarrow}\xi^{\mu(2)} - g^{\mu(2)}\Big[ \frac{s-2}{d}\dd - 2\tanh z \Big]\xi\;.
\end{equation}
The other gauge transformations \eqref{e:gauge_L} ($n\neq 2$) remain intact. Note that \gt has lost its unified form \eqref{e:gauge_L} for all $n$. The last term in \eqref{e:gauge_L2} is not the lowering or raising operator. So, the argument that action must contain only interaction between the spin--$0,1,2$ fields in the spin--$0$ sector is not valid anymore. This is the reason behind the appearance of the non--vanishing $S_{30}$ term.

Now, the action \eqref{e:ansatz_nn}--\eqref{e:ansatz_nn-2} (except for $S_{20}$, $S_{10}$, $S_{00}$) with the new terms \eqref{e:act2_a}--\eqref{e:act2_d} along with \gt \eqref{e:gauge_L}, \eqref{e:gauge_L2} does not have factors with the pole $d-2$. 

\section{Eliminating the slicing coordinate} \label{s:elim}

Since our aim is to describe the \1ads theory in terms of exclusively objects in \0ads, the slicing coordinate $z$ has to be eliminated. To do this the Kaluza--Klein mode expansion is introduced. The component fields and gauge parameters are to be expanded on the complete set of basis functions $P^{n}_k(z)$. Then, the $z$--dependence is integrated out from most of the component actions \eqref{e:ansatz_nn}--\eqref{e:ansatz_nn-2}, \eqref{e:act2_b}, \eqref{e:act2_d} and the gauge transformations \eqref{e:gauge_L}, \eqref{e:gauge_L2} via the basis functions and their properties. In order to get rid of the remaining $z$--dependence, the gauge is partially fixed. All but two spin--0 and spin--1 Stueckelberg modes are set to zero vanishing the remaining terms in the component actions \eqref{e:act2_a}, \eqref{e:act2_c}. 

The basis functions are specifically chosen to integrate out the slicing coordinate most effectively. So, individual basis functions are used for different component fields. Except for the spin--0 sector which will be discussed below, the basis functions $P^{n}_k$ in each spin--$n$ sector are the eigenfunctions of the second--order differential operator $L_{d+2n-4}L_\downarrow$ in \eqref{e:ansatz_nn}\footnote{For a discussion on the second--order differential operator and its eigenfunctions see \bref{s:Jac}.}
\begin{all}\label{e:mod_exp_f}
  \phi^{\mu(n)}(x,z) &= \sum^{\infty}_{k=0}\phi^{\mu(n)}_k(x)P^n_k(z)\;,&& 1\leqslant n \leqslant s\;,\\
  \phi(x,z) &= \sum^{\infty}_{k=0}\phi_k(x)P^1_k(z)\;,&&n = 0\;,
\end{all}
where $P^n_k$ are constructed from the Jacobi polynomials \eqref{basis_f}. The set of the basis functions $\{P^n_k\}^\infty_{k=0}$ for a given level $n$ is orthonormal \eqref{e:inn_prod_PP} with respect to integration with the weight function $(\cosh z)^{d+2n-2}$ \eqref{e:inn_prod_PP} which coincides with the one in \eqref{e:ansatz_nn} (keep in mind the hidden integration measure in the double--integral notation). Also, the neighbouring basis functions $P^n_k$ and $P^{n+1}_{k-1}$ have the relation via the lowering/raising operators \eqref{e:L_down}, \eqref{e:L_up}. Thus, after applying the $L$--operators in \eqref{e:ansatz_nn}--\eqref{e:ansatz_nn-2} (except for spin--0 sector which will be treated separately) only the orthonormal property has to be used. All these features make the basis functions $\{P^n_k\}^\infty_{k=0}$ the most advantageous choice for the mode expansion. 

As already discussed in \cite{Alkalaev:2021zda}, despite the fact that it would be natural to use $P^0_k(z)$ for the spin--$0$ component field these functions possess the orthonormal property only for $d>2$ which makes them unsuitable for our consideration. Furthermore, the use of $P^0_k(z)$ is not necessary since the action has lost its elegant structure of the lowering/raising operators in the spin--0 sector \eqref{e:act2_a}--\eqref{e:act2_d}. That is why we exploit the same functions $P^1_k(z)$ for the both spin--$1$ and spin--$0$ fields. 

The component gauge parameters need to be selected according to the component field expansion
\begin{all}\label{e:mod_exp_gp}
  \xi^{\mu(n)}(x,z) &= \sum^{\infty}_{k=0}\xi^{\mu(n)}_k(x)P^{n+1}_k(z)\;,&& 1\leqslant n \leqslant s\;,\\
  \xi(x,z) &= \sum^{\infty}_{k=0}\phi_k(x)P^2_k(z)\;,&& n=0\;.
\end{all}
Note that the basis function $P^{n}_k(z)$ of the component gauge parameter of rank $(n-1)$ is purposefully one level higher than the one of the component field of rank $n$. The reason is that the level $n$ of the basis functions of all three terms in the gauge transformation \eqref{e:gauge_L} after applying $L$--operators on the right--hand side coincide with the basis function level on the left--hand side (except for the spin--2 sector \eqref{e:gauge_L2}).  

Having the mode expansions \eqref{e:mod_exp_f}, \eqref{e:mod_exp_gp}, one can integrate out the slicing coordinate $z$ in most sectors of the gauge transformation \eqref{e:gauge_L}, \eqref{e:gauge_L2}
\begin{all}\label{e:gauge_mod_exp}
  &\delta \phi^{\mu(2)}_k = \nabla^{\mu}\xi^{\mu}_k + (s-2)\gamma_{3|k-1}\xi^{\mu(2)}_{k-1} -g^{\mu(2)}\sum^{\infty}_{l=0}\Big( P^2_k,\Big[\frac{s-2}{d}\dd - 2\tanh z \Big]P^1_l \Big)_2\xi_l  \;,\hphantom{aaaaaa} n = 2\;,\\
  &\delta \phi^{\mu(n)}_k = \nabla^{\mu}\xi^{\mu(n-1)}_k + (s-n)\gamma_{n+1|k-1}\xi^{\mu(n)}_{k-1} - \frac{2(d+n+s-4)}{(d+2n-6)(d+2n-4)} \gamma_{n|k} g^{\mu(2)}\xi^{\mu(n-2)}_{k+1} \;, n\neq 2\;,
\end{all}
where $\gamma_{n|k}$ are constants resulting from applying the $L$--operators on the basis functions \eqref{e:gamma} and $(\hphantom{a},\hphantom{a})_2$ is an inner product between basis functions defined in \eqref{e:inn_prod}. 

As explained above, using only the orthonormal condition and the lowering/raising property of the $L$--operators the slicing coordinate can be eliminated from the component actions 
\begin{all}\label{e:act_mod_exp}
  S_{nn} =  &\int\sum^{\infty}_{k=0} A_n\Big[ -(\nabla_\mu \phi^{\nu(n)}_k)^2 + n(\nabla_\mu\phi^{\mu\nu(n-1)}_k)^2 - n(n-1)\nabla_\mu \phi'_{k\,\nu(n-2)}\nabla_\rho \phi^{\mu\nu(n-2)\rho}_k +\\
  & +\frac{n(n-1)}{2}(\nabla_\mu \phi'^{\nu(n-2)}_k)^2 + \frac{n(n-1)(n-2)}{4}(\nabla_\mu\phi'^{\mu\nu(n-3)}_k)^2   \Big] +\\
  &+ \Big( B_n -(\gamma_{n|k})^2D_n \Big) (\phi^{\mu(n)}_k)^2  + \Big( C_n  -(\gamma_{n|k})^2E_n \Big)(\phi'^{\mu(n-2)}_k)^2 \;, \\
  S_{nn-1} = &\int\sum^{\infty}_{k=0}-\gamma_{n|k}\Big[ F_n\nabla_\mu\phi^{\mu\nu(n-1)}_k\phi_{k+1\,\nu(n-1)} + G_n\nabla_\mu\phi'_{k\,\nu(n-2)}\phi^{\mu\nu(n-2)}_{k+1} + \\
  &+  H_n\nabla_\mu\phi'^{\mu\nu(n-3)}_k\phi'_{k+1\,\nu(n-3)}\Big]\;,\\
  S_{nn-2} = &\int\sum^{\infty}_{k=0}\gamma_{n|k}\gamma_{n-1|k+1} I_n\phi'^{\mu(n-2)}_k\phi_{k+2\,\mu(n-2)}\;,
\end{all}
for all $0\leqslant n \leqslant s$ except for $S_{30}$, $S_{20}$, $S_{10}$, $S_{00}$. The latter are to be treated separately. 

Same steps applied to the modified component actions $S_{30}$, $S_{20}$, $S_{10}$, $S_{00}$ lead to double sums $\sum^{\infty}_{k,l=0}$ with factors composed of inner products $(\hphantom{a},\hphantom{a})_n$ between the basis functions of different levels. Fortunately, we need not evaluate all of them because the component gauge parameters $\xi_{k\geqslant 0}$ and $\xi^{\mu}_{k\geqslant 0}$ can be exhausted to set to zero the spin--$0$ and spin--$1$ modes $\phi_{k\geqslant 1}\rightarrow 0$ and $\phi^{\mu}_{k\geqslant 1}\rightarrow 0$ for they are the Stueckelberg fields \eqref{e:gauge_mod_exp}
\begin{equation}
    \delta\phi_{k+1} = s\gamma_{1|k}\xi_{k}\;, \qquad \delta\phi^{\mu}_{k+1} = (s-1) \gamma_{2|k}\xi_{k}^{\mu} \;.
\end{equation}

Vanishing the Stueckelberg spin--0,1 fields drastically simplifies the action. Let us denote the spin--2, spin--1 and spin--0 fields in a more convenient way $h^{\mu\nu} = \phi^{\mu\nu}_0$, $h = \phi'_0$, $A^\mu = \phi^\mu_0$, $\varphi = \phi_0$ so we can easier distinguish between them later. Then, the remaining component actions read  
\begin{al}
  S_{30} & =0   \;,  &S_{20} & = \int \Big[ W \big(\nabla_\mu\nabla_\nu h^{\mu\nu} - \Box h \big) +  X h\Big]\varphi    \;, \label{e:S_20} \\
  S_{10} & = 0  \;, & S_{00} & = \int \Big\{ Y (\nabla_\mu \varphi)^2 + Z \varphi^2 \Big\}\;,\label{e:S_00}
\end{al}
where the constants $W,\ldots,Z$ are known constants, see Appendix \bref{ap:An-In}. The main reason why most of the terms vanish is because the raising operator annihilates the lowest mode basis function $L_d P^1_0 = 0$, see \eqref{e:L_up}. Note that for the same reason the component action $S_{31}$ is also zero. 

To conclude, we systematically excluded the $z$--dependence by means of the partial gauge fixing. It should be pointed out, that sectors of the spin--1 mode--0 field $A^\mu$, and the spin--0,2 mode--0 fields $\varphi$, $h^{\mu(2)}$, and all the other fields $\phi^{\mu(2)}_{k\geqslant 1},\phi^{\mu(3)}_{k\geqslant 0},\ldots,\phi^{\mu(s)}_{k\geqslant 0}$ propagate separately and do not interact with each other. In next Section \bref{s:EOM} the equations of motion of each of these three field sectors are considered separately.

\section{Equations of motion}\label{s:EOM}

For the sake of generality, in the previous Sections the dimension $d$ was held arbitrary. Now, to take the final step of the \3ads/\2ads degression and obtain the resulting theory spectrum, we need to restrict ourselves to $d=2$. In this dimension one can take advantage of the so--called Schouten identities \cite{Conde:2016izb,Mkrtchyan:2017ixk,Kessel:2018ugi}. Such expressions can remove most of the kinetic terms in the EOM of fields of spin higher than 2 which is helpful to analyze the spectrum.  

These identities exploit the fact that a product three Kronecker deltas with antisymmetrized indices in $d=2$ identically vanishes $\delta^{[\mu}_\alpha\,\delta^\nu_\beta\,\delta^{\rho ]}_\gamma \equiv 0$. Contracting it with a second derivative of a totally--symmetric tensor field $X^{\mu(n)}$ with at least two indices $n\geqslant 2$ yield the following identity
\begin{equation}
    \delta^{[\mu}_\alpha\,\delta^\nu_\beta\,\delta^{\rho ]}_\gamma \,\nabla_\nu\nabla^\beta {X_\rho}^{\gamma\lambda(n-2)} \equiv 0\;.
\end{equation}
Then, raising index $\alpha$ and symmetrizing free indices, the Schouten identity can be expressed as
\begin{all}\label{Schouten}
   &\Box X^{\mu(n)} - \frac{2}{n}\nabla^{\mu}\nabla_\nu X^{\mu(n - 1)\nu} +\frac{1}{n(n-1)}\nabla^{\mu}\nabla^{\mu}X'^{\mu(n - 2)} + \frac{2}{n(n-1)}g^{\mu(2)}\big[ \nabla_\nu\nabla_\rho X^{\mu(n - 2)\nu\rho} - \\
  &-\Box X'^{\mu(n - 2)}\big] +n X^{\mu(n)} - \frac{2}{n} g^{\mu(2)}X'^{\mu(n - 2)} \equiv 0\;,
\end{all}
where the algebraic terms arise from commutating covariant derivatives, see \eqref{e:commutator}. In most cases the Schouten identity will be used for TT gauged fields for which it takes the simplified form $\Box X^{\mu(n)} \equiv -n X^{\mu(n)} $. 

In what follows we consider each of the three field sectors separately since they are independent of each other. Recall that the spin--0,1 mode--$k$ fields ($k\geqslant 0$) are already gauge fixed to zero. 

\subsection{Spin--1 sector}

This field sector contains only one field. The EOM for the spin--1 mode--0 field from the action \eqref{e:act_mod_exp} read 
\begin{equation}\label{}
     A_1\big[ \Box A^{\mu} - \nabla^{\mu}\nabla_\nu A^\nu\big]+ \big( B_1 - (\gamma_{1|0})^2D_1\big)A^{\mu} = 0\;,
\end{equation}
It describes the Proca equation so that there is the Lorentz condition which arises as a differential consequence. Inserting the known coefficients \eqref{e:A_n}, \eqref{Bn}, \eqref{Dn}, \eqref{e:gamma} into , we obtain
\begin{all}
  &\Big[ \Box -\big(s(s-1)-1\big)\big]A^{\mu} = 0\;,\\
  &\nabla_\mu A^\mu = 0\;.
\end{all}
The spin--1 field is the first non--trivial constituent of the theory spectrum. The mass parameter coincides with the one prescribed by the group--theoretical analysis in \cite{Alkalaev:2021zda}. 

\subsection{Spin-0 and spin--2 sector}

For the $n=2$ case the Schouten derivative terms exactly coincide with the kinetic operator in EOM of the component spin--2 field $h^{\mu\nu}$. So, the EOM in the spin--2 sector is simplified. As a result, the spin--2 equation from a differential equation turns into an algebraic equation on the spin--2 field. The spin--0 sector remains intact for the Schouten identities are formulated only for fields of spin 2 and higher. Thus, the spin--2,0 mode--0 EOM from \eqref{e:act_mod_exp}, \eqref{e:S_20}, \eqref{e:S_00} read
\begin{al}
  & \mathcal{A}_{2|0} h^{\mu\nu} + \mathcal{B}_{2|0} g^{\mu\nu}h + W\nabla^\mu\nabla^\nu\varphi - Wg^{\mu\nu}\Box\varphi + Xg^{\mu\nu}\varphi = 0\;,\label{e:5}        \\
  & - Y\Box\varphi + 2Z\varphi + W\nabla_\mu\nabla_\nu h^{\mu\nu} - W\Box h + Xh  = 0\;,\label{e:6}
\end{al}
where for later convenience we define the following coefficients
\begin{equation}
    \mathcal{A}_{n|k} = 2\big(-nA_n + B_n - (\gamma_{n|k})^2D_n\big)\;,\qquad \mathcal{B}_{n|k} = 2\big( nA_n + C_n - (\gamma_{n|k})^2E_n \big)\;,
\end{equation}
which are constructed from the known constants, see \eqref{e:A_n}--\eqref{En}, \eqref{e:gamma}. 

In order to simplify the calculations below, let us redefine the scalar field $\sqrt{\frac{2}{3}}\varphi \rightarrow \varphi$. The first equation \eqref{e:5} allows us to express the spin--2 mode--0 in terms of the scalar field
\begin{all}
  &h = \frac{1}{s(s-1)}\Box \varphi - \frac{3s-2}{s}\varphi\;,\\
  &h^{\mu\nu} = \frac{1}{s(s-1)}\nabla^\mu\nabla^\nu\varphi - \frac{3s-2}{2s}g^{\mu\nu}\varphi\;,
\end{all}
where the denominator never vanishes since the equation takes place only for $s\geqslant 2$. Thus, the spin--2 mode--0 field is auxiliary and does not contribute to the spectrum. Substituting these expressions into the last equation \eqref{e:6}, we are left with the fourth-order equation for the spin--0 mode--0 field only
\begin{all}
  &2\sqrt{\frac{2}{3}}\big[ -Y\Box + Z \big] \varphi + W\Big[ \frac{1}{s(s-1)}\nabla_\mu\nabla_\nu\nabla^\mu\nabla^\nu -\frac{3s-2}{2s}\Box - \frac{1}{s(s-1)}\Box\Box +  \frac{3s-2}{s}\Box \Big]\varphi +\\
  &+X\Big[ \frac{1}{s(s-1)}\Box - \frac{3s-2}{s}\Big]\varphi = 0
\end{all}
This equation is in fact second--order because all fourth--order terms cancel each other. So the scalar field is actually subject to the Klein-Gordon equation
\begin{equation}
  \Big[\Box - s(s - 1)\Big]\varphi = 0\;,
\end{equation}
where the mass parameter again coincides with the expected value from \cite{Alkalaev:2021zda}.

\subsection{Higher--spin sector}\label{s:hs}

Now, it is left to show that the remaining higher--spin fields do not contribute to the spectrum. Roughly speaking, we impose the TT gauge. Then, all equations take the Klein--Gordon form $(\Box + m^2) \phi^{\mu(n)}=0$ with some mass parameter $m^2$. Using the Schouten identities, we substitute the second--derivative term with an algebraic one which implies that $m'^2 \phi^{\mu(n)}=0$ where $m'^2$ is a new parameter. 

However, this line of reasoning is not complete. Firstly, the tracelessness and transversality conditions can not be imposed simultaneously on an arbitrary field since two differential equations on a gauge parameter with the field as a source lead to a consistency condition of the form $\nabla_{\nu}\nabla_{\rho}\phi^{\mu(n-2)\nu\rho}_k + \ldots = 0$ where dots represent trace terms. Such a consistency condition can not be satisfied in general. Instead, one can impose only tracelessness, check the consistency condition $\nabla_{\nu}\nabla_{\rho}\phi^{\mu(n-2)\nu\rho}_k = 0$ (double transversality) on--shell, and then gauge fix transversality condition. Secondly, the resulting constant $m'^2$ can turn out to be zero in which case the EOM is satisfied identically and therefore does not determine the field.

Let us outline the general strategy. The following steps are demonstrated on the first non-trivial and instructive case of the spin $s=4$ in Appendix \bref{s:s4}.
\begin{enumerate}
    \item Via the left--over gauge transformation \eqref{e:gauge_mod_exp} with the gauge parameters $\xi^{\mu(n)}_k$ ($2\leqslant n\leqslant s-1$, $k\geqslant 0$) the tracelessness condition on the fields $\phi^{\mu(n)}_k$ ($3\leqslant n\leqslant s$, $k\geqslant 0$) is imposed
    \begin{equation}
        \phi'^{\mu(n-2)}_k = 0\;,\quad 3\leqslant n\leqslant s\;,\;\;k\geqslant 0\;.
    \end{equation}
    This gauge fixing is consistent since the number of the gauge parameters are exactly the same as the conditions on them. The spin--2 field $\phi^{\mu(2)}_k$ ($k\geqslant 1$) remains intact.
    
    Note that the higher--spin sector of the EOM from \eqref{e:act_mod_exp} is divided into independent subsectors with the following fields $\phi^{\mu(2)}_k, \phi^{\mu(3)}_{k-1}, \ldots,\phi^{\mu(n)}_{k-n+2}$ ($k\geqslant 1$), where the series ends when  $k-n+2=0$ or $n=s$. The consideration below remains the same in each subsector. 
    
    \item We start analyzing each subsector from the EOM on the lowest--rank field, i.e. the spin--2 field, and iteratively move to the EOM on the highest--spin field, i.e. either the spin--$(k+2)$ or spin--$s$ field. The spin--2 EOM with the applied Schouten identity and traceless gauge reads
    \begin{equation}\label{n2}
        \mathcal{A}_{2|k}\phi^{\mu(2)}_k + \mathcal{B}_{2|k}g^{\mu(2)}\phi'_k - \gamma_{3|k-1}F_3\nabla_\nu \phi^{\mu(2)\nu}_{k-1} = 0\;.
    \end{equation}
    Contracting equation \eqref{n2} with the metric, we obtain the equation on the trace of the spin--2 field only
        \begin{equation}
        (\mathcal{A}_{2|k} + 2\mathcal{B}_{2|k})\phi'_k  = 0\;.
    \end{equation}
    The factor $\mathcal{A}_{2|k} + 2\mathcal{B}_{2|k}\neq 0$ for all $s\geqslant 2$ and $k\geqslant 1$ so the trace can be uniquely determined $\phi'_k=0$. Now, the trace is substituted back into equation \eqref{n2} 
        \begin{equation} \label{n23}
        \phi^{\mu(2)}_k = \frac{\gamma_{3|k-1}F_3}{\mathcal{A}_{2|k}} \nabla_\nu \phi^{\mu(2)\nu}_{k-1}\;,
    \end{equation}
    here $\mathcal{A}_{2|k} \neq 0$ for all $s\geqslant 2$ and $k\geqslant 1$. Thus, the spin--2 field is fully determined in terms of the divergence of the spin--3 field $\nabla_\nu \phi^{\mu(2)\nu}_{k-1}$. 

\item  After substituting the equation \eqref{n23} into spin--3 EOM and taking the trace, one can show that for all $s\geqslant 3$ and $k\geqslant 1$ 
\begin{equation}
    \nabla_\nu\nabla_\rho \phi^{\mu\nu\rho}_{k-1} = 0\;,
\end{equation}
which coincides with the consistency condition that allows to impose the transversality gauge on the traceless field. Thus, we proved that the TT gauge can eventually be imposed on the spin--3 field on--shell. In the sequel, the spin--3 field is considered to be traceless $\phi'^{\mu}_{k-1}=0$ and transversal $\nabla_\mu \phi^{\mu\nu(2)}_{k-1} = 0$. 

\item Since the spin--3 field is TT gauge fixed, the equation \eqref{n23} yields $\phi^{\mu(2)}_k = 0$ ($k\geqslant 1$). Hence, the spin--2 field is non--dynamical. 

Now, the spin--3 EOM (after applying the spin--3 Schouten identity) has the form 
\begin{equation}
    \mathcal{A}_{3|k-1}\phi^{\mu(3)}_{k-1} - \gamma_{4|k-2}F_4\nabla_{\nu}\phi^{\mu(3)\nu}_{k-2} = 0\;.
\end{equation}
If $\mathcal{A}_{3|k-1} \neq 0$, the consequence of the spin--3 transversality is the spin--4 double transversality $\nabla_{\nu}\nabla_\rho\phi^{\mu(2)\nu\rho}_{k-2} = 0$. Hence, the spin--4 field can also be TT gauge fixed. The case $\mathcal{A}_{3|k-1} = 0$ will be considered in the next step.

\item The previous step is to be recurrently repeated until the last but one field. Suppose the spin--$l$ field is considered ($3\leqslant l\leqslant n - 1$). From the previous step this spin--$l$ field is TT gauge fixed which means that the previous spin--$(l-1)$ field is zero and the next spin--$(l+1)$ field is double--transversal because of the EOM
\begin{equation}\label{s_l_eom}
    \mathcal{A}_{l|k-l+2}\phi^{\mu(l)}_{k-l+2} - \gamma_{l+1|k-l+1}F_{l+1}\nabla_{\nu}\phi^{\mu(l)\nu}_{k-l+1} = 0\;.
\end{equation}
The double transversality of the spin--$(l+1)$ field leads to the TT gauge in the next step.

Suppose $\mathcal{A}_{l|k-l+2} = 0$.\footnote{E.g. it happens for the following values: $s=12$, $l=10$, $k-l+2 = 5$. } Then, on the one hand the spin--$l$ EOM \eqref{s_l_eom} does not determine the spin--$l$ field. On the other hand it simply tells that the spin--$(l+1)$ field is already transversal. So, it need not to be TT gauge fixed via the spin--$l$ gauge parameter and thus the spin--$l$ gauge parameter can now be used to remove the spin--$l$ field as the Stueckelberg field
\begin{equation}
    \delta \phi^{\mu(l)}_{k-l+2} = \xi^{\mu(l)}_{k-l+1}\;.
\end{equation}
This Stueckelberg gauge fixing is consistent since the field and the parameter are both in the TT gauge.\footnote{The gauge parameter is traceless by the definition. After the gauge fixing of a trace at step 1 there is the left--over gauge condition of the form $\nabla_{\nu}\xi^{\mu(l-1)\nu}_{k-l+1} + \xi^{\mu(l-1)}_{k-l+2} = 0$. The spin--$(l-1)$ gauge parameter has already been exhausted at the previous step. So, the resulting left--over gauge condition is the transversality of the spin--$l$ gauge parameter. }

\item The final step is to prove that the last equation (either $n=s$ or $k-n+2=0$) 
\begin{equation}
    \mathcal{A}_{n|k-n+2}\phi^{\mu(n)}_{k-n+2} = 0\;.
\end{equation}
always has a trivial solution. Indeed, one can straightforwardly demonstrate that $\mathcal{A}_{n|k-n+2} \neq 0$ for $n=s$ or $k-n+2=0$ and hence the highest--spin field in the subsector vanishes $\phi^{\mu(n)}_{k-n+2} = 0$.

\end{enumerate}

Thus, the higher--spin sector does not contain dynamical fields. All the fields after solving the EOM and proper gauge fixing vanish. 

To conclude, the Kaluza--Klein spectrum of the \3ads Fronsdal theory of the spin--$s$ massless field seen from the  \2ads perspective coincide with that one predicted from the group--theoretical approach elaborated in \cite{Alkalaev:2021zda}. Instead of an infinite tower of Kaluza--Klein modes it consists of only two fields: the massive Klein--Gordon and Proca fields of equal energies $E=s$. The other modes are Stueckelberg, auxiliary, or vanishing on--shell.   


%


\paragraph{Acknowledgements.} I thank K. Alkalaev for his encouragement of the current
work. Also, I am grateful to all LPI QFT seminar participants, especially to A. Smirnov and M. Vasiliev, for feedback on a presentation of some of the results contained in this paper. This work was supported by the
Foundation for the Advancement of Theoretical Physics and Mathematics “BASIS”.

\appendix

\section{Conventions and notation}\label{a:Con_not}


\paragraph{\0ads and \1ads geometry.} Firstly, we introduce the index notation. Greek indices correspond to \0ads geometry $\mu,\nu = (0,\ldots,d-1)$  and Latin indices refer to \1ads geometry $m,n = (0,\ldots,d-1,\bullet)$.  

The degression means that objects (such as fields, gauge parameters, the metric) in \1ads are to be rewritten in terms of \0ads geometry. The natural way to do so is to use the following form of the metric \eqref{metric}
\begin{equation}\label{e:gbar}
\ol g_{mn}(x,z)=
\left(
\begin{array}{c|c}
\cosh^2 z\, g_{\mu\nu}(x) & 0 \\
\hline
0 & 1
\end{array}
\right)\;.
\end{equation}
This equation implies that the \1ads metric is foliated into \0ads slices with the slicing coordinate $z\in (-\infty, \infty)$. A bar over the metric, covariant derivative, Riemann tensor and Christoffel symbols refers to \1ads; the mentioned objects without a bar refer to \0ads. Using \eqref{e:gbar}, it is straightforward to derive expressions for the Riemann tensors and Christoffel symbols in both spaces
\begin{equation}
\ol R_{mnkl} = - \ol g_{mk}\ol g_{nl} + \ol g_{ml}\ol g_{nk}\;,\quad  R_{\mu\nu\rho\lambda} =  - g_{\mu \rho} g_{\nu \lambda} + g_{\mu \lambda} g_{\nu \rho}\;,
\end{equation}
\begin{equation}
\overline{\Gamma}^\rho_{\mu\nu} = \Gamma^\rho_{\mu\nu}\;,\quad \overline{\Gamma}^\mu_{\nu\bullet} = \tanh z\,\delta^\mu_\nu\;, \quad \overline{\Gamma}^{\bullet}_{\mu\nu} = - \cosh^2 z\tanh z\,g_{\mu\nu}\;.
\end{equation}

A commutator of covariant derivatives is expressed in terms of the Riemman tensors as follows
 \begin{equation}\label{e:commutator}
 \left[  \nabla_\mu, \nabla_\nu  \right] {X^{\alpha\cdots}}_{\beta\cdots} =  R_{\mu\nu\phantom{\alpha}\rho}^{\phantom{\mu\nu}\alpha}{X^{\rho\cdots}}_{\beta\cdots} +\ldots
 + R_{\mu\nu\beta}^{\phantom{\mu\nu\beta}\rho}{X^{\alpha\cdots}}_{\rho\cdots}+\ldots\;,
 \end{equation}

\paragraph{Symmetrization.} 
Since we are dealing with tensors with multiple indices, it is natural to introduce a more elegant notation of tensorial structures originally introduced in \cite{Vasiliev:1986td}.  A field with $n$ symmetric indices is notated as
\begin{equation}\label{sym_rule}
    \phi^{\mu_1\mu_2\ldots\mu_n} \equiv  \phi^{\mu(n)}\;.
\end{equation}
Symmetrization is denoted as repeating indices and is understood as a sum of a minimal number of terms without a weight. So, e.g. the symmetrization of the covariant derivative and the totally--symmetric gauge parameter reads as follows
\begin{equation}
    \nabla^\mu \xi^{\mu(n)} \equiv \nabla^{\mu_1} \xi^{\mu_2\ldots\mu_{n+1}} + \text{cycle}\;.
\end{equation}
Note that $\nabla^\mu\nabla^\mu h$, where $h$ is a scalar, contains only one term for it is symmetric already.

Antisymmetrization is denoted as square bracket $[\ldots]$. The overall factor does not matter for our considerations. 

\paragraph{Action notation.}

Squared terms in the action mean the contraction of an expression within the parenthesis with itself e.g.
\begin{equation}
    (\nabla_\mu \phi^{\nu(n)})^2 = \nabla_\mu \phi^{\nu(n)} \nabla^\mu \phi_{\nu(n)}\;.
\end{equation}

The integration measure read
\begin{all}\label{double_int}
    &\int d\mu_{d+1} = \int d^dx\int dz \sqrt{-\ol g}  = \int d^dx \sqrt{-g}\int dz \cosh^dz \equiv \iint\;,\\
    &\int d\mu_{d} = \int d^dx \sqrt{- g} \equiv \int\;,
\end{all}
where $\ol g = \det \ol g_{mn}$, $g = \det g_{\mu\nu}$.

\section{Jacobi polynomials}\label{s:Jac}

In this Appendix we review the basics concerning the Jacobi polynomials\footnote{For a more comprehensive description of the Jacobi polynomials see e.g. \cite{Shen2011}.} required to construct the basis functions. The Jacobi polynomials can be represented via the hypergeometric function
\begin{equation}
J^{\alpha,\beta}_k(x)  = \frac{\Gamma(k+\alpha+1)}{k ! \Gamma(\alpha+1)} {}_2F_1(-k,k+\alpha+\beta+1;\alpha+1;\frac{1-x}{2})\,,
\end{equation}
where $\alpha,\beta > -1$, and $k \in  \mathbb{N}_0$, and $x\in (-1,1)$. These polynomials are the eigenfunctions of the following Sturm--Liouville operator
\begin{equation}\label{e:stu-lio}
 -(1-x)^{-\alpha}(1+x)^{-\beta} \partial_{x}\left[(1-x)^{\alpha+1}(1+x)^{\beta+1} \partial_{x} J^{\alpha,\beta}_k(x)\right] =\lambda_{k}^{\alpha, \beta} J_{k}^{\alpha, \beta}(x)\;,
\end{equation}
with the eigenvalues $\lambda_{k}^{\alpha, \beta}=k(k+\alpha+\beta+1)$. Also, for given $\alpha,\beta$ they form an orthogonal basis in the $\omega^{\alpha,\beta}$--weighted $L^2(\mathbb{R})$ space
\begin{equation}\label{e:inn_prod_J}
\int_{-1}^{1} J_{k}^{\alpha, \beta}(x) J_{l}^{\alpha, \beta}(x) \omega^{\alpha, \beta}(x) d x=\left\|J_{k}^{\alpha, \beta}\right\|^2 \delta_{m n}\,,
\end{equation}
where the weight function is $\omega^{\alpha,\beta} = (1-x)^\alpha(1+x)^\beta$ and the normalization constant reads 
\begin{equation}\label{e:inn_prod_J_norm}
\left\|J_{k}^{\alpha, \beta}\right\|^2=\frac{2^{\alpha+\beta+1} \Gamma(k+\alpha+1) \Gamma(k+\beta+1)}{(2 k+\alpha+\beta+1) k ! \Gamma(k+\alpha+\beta+1)}\,.
\end{equation}
Note that \eqref{e:stu-lio} as a second--order differential equation contains the second solution. But the other solution does not form a normalizable basis with respect to this inner product \eqref{e:inn_prod_J_norm} which does not suit our consideration.

The Jacobi polynomials possess a powerful property that $J_{k}^{\alpha, \beta}$ and $J_{k-1}^{\alpha+1, \beta+1}$ are related through the derivative
\begin{equation} \label{e:dJ}
\partial_x J^{\alpha,\beta}_{k}(x) = \mu^{\alpha,\beta}_{k}J^{\alpha+1,\beta+1}_{k-1}(x)\,,\quad \mu^{\alpha,\beta}_{k}=\frac{1}{2}(k+\alpha+\beta+1)\,.
\end{equation}

\paragraph{Basis functions.} Before introducing basis functions itself, let us define the so--called raising and lowering operators, or the $L$--operators
\begin{all} \label{def_L}
&L_{m}A(z) \equiv \cosh^{-m} z \partial (\cosh^m z A) \equiv (\partial+m\tanh z)A\,,\\
&L_{\downarrow}A(z) \equiv \cosh^2z\,L_2A\;,
\end{all}
where $m = d+2n-4$ and for the sake of simplifying the notation the $z$ derivative denoted as $\dd$. 

Now, assume that the parameters in the Sturm--Liouville equation coincide and are equal to $\alpha = \beta = (d+2n-4)/2$.  Substitution of the variable $x = -\tanh z$ in the differential equation \eqref{e:stu-lio} provides us with\footnote{Such a factorization of the second--order differential equations on the lowering/raising operators is considered in \cite{RevModPhys.23.21}.}
\begin{equation}\label{e:LLgamma}
    L_{d+2 n-4}L_{\downarrow} P^{n}_{k}(z)  =-\left(\gamma_{n|k}\right)^{2} P_{k}^{n}(z)\;,
\end{equation}
where the eigenvalue is 
\begin{equation} \label{e:gamma}
\gamma_{n|k} =\sqrt{(k+1)(k+d+2 n-4)}\,.
\end{equation}
and $P^{n}_{k}(z) $ are proportional to the original Jacobi polynomials  
\begin{equation}\label{basis_f}
    P^n_k(z) = N^n_k(\cosh z)^{-d-2n+2} {}_2F_1(-k,k+d+2n-3;\frac{d+2n-2}{2};\frac{1+\tanh z}{2})\,.
\end{equation}
The functions $P^n_k(z)$ are called the basis functions of level $n$. The normalization constant here is chosen in a way that $P^{n}_{k}(z) $ form an orthonormal basis in the $(\cosh z)^{d+2n-2}$--weighted $L^2(\mathbb{R})$ space
\begin{equation}\label{e:Norm}
    N^n_k =  \frac{\sqrt{d+2n-3+2k}}{2^{\frac{d+2n-3}{2}}\Gamma(\frac{d+2n-2}{2})}\sqrt{\frac{\Gamma(d+2n-3+k)}{k!}}\,.
\end{equation}

The inner product of level $n$ between two functions $A(z)$ and $B(z)$ is defined as
\begin{equation}\label{e:inn_prod}
    (A,B)_n \equiv \int^{+\infty}_{-\infty}dz(\cosh z)^{d+2n-2} A(z) B(z)  \,.
\end{equation}
Then, according to the orthogonal property of the Jacobi polynomials \eqref{e:inn_prod_J}, \eqref{e:inn_prod_J_norm} and the choice of the normalization condition \eqref{e:Norm}, the inner product of level $n$ between two basis functions of the same level are  
\begin{equation}\label{e:inn_prod_PP}
    (P^n_k,P^n_l)_n \equiv \int^{+\infty}_{-\infty}dz(\cosh z)^{d+2n-2} P^n_k(z) P^n_l(z)  =\delta_{kl}\,.
\end{equation}

After expressing the equation \eqref{e:dJ} in terms of a new variable $z$ and the basis functions, one yields 
\begin{equation}\label{e:L_up}
    L_{d+2 n-2} P^{n}_{k}=-\gamma_{n+1|k-1} P^{n+1}_{k-1}\,,
\end{equation}
i.e. the operator $L_{d+2 n-2}$ transfers a basis function from level $n$ to a higher level $n+1$. That is the reason for calling it the raising operator. Applying this expression to the equation \eqref{e:LLgamma}, one can prove that 
\begin{equation}\label{e:L_down}
    L_{\downarrow} P^{n}_{k} = \gamma_{n|k} P^{n-1}_{k+1}\;.
\end{equation}
Thus, the operator $L_{\downarrow}$ does the opposite. It decreases the level of a basis function. Moreover, the both operators can transform into one another within the inner product 
\begin{equation}
    (A,L_{\downarrow}B)_n = -(L_{d+2n-2}A,B)_{n+1}\,.
\end{equation}

\section{Constructing the ansatz action}\label{s:how}

Properties of the raising and lowering operators to transfer the basis functions between levels have far--reaching implications. In this Appendix, we give some comments on how the ansatz action \eqref{e:ansatz_nn}--\eqref{e:ansatz_nn-2} can be guessed.

In Section \bref{s:elim} we attributed the basis functions of level $n$ to the rank--$n$ field $\phi^{\mu(n)}$ and the rank--$(n-1)$ gauge parameter $\xi^{\mu(n-1)}$ as shown on the following diagram\footnote{The case of the scalar field ($n=0$) in two dimensions is exceptional as explained in Section \bref{s:elim} and will not be considered in this Appendix.}
\begin{equation}
  \begin{tikzcd}
    \phi^{\mu(n + 1)} & P^{n + 1} \ar[d,shift left=1ex,"L_{\downarrow}"] & \xi^{\mu(n)}\\
    \phi^{\mu(n)} & P^{n} \ar[u,shift left=1ex,"L_{d+2n - 2}"] \ar[d,shift left=1ex,"L_{\downarrow}"] & \xi^{\mu(n - 1)}\\
    \phi^{\mu(n - 1)} & P^{n - 1} \ar[u,shift left=1ex,"L_{d+2n - 4}"]\, & \xi^{\mu(n - 2)}
  \end{tikzcd}
\end{equation}
For example, applying the raising $L$--operators two times to the rank--$(n-2)$ field after the mode expansion one obtains the rank--$(n-2)$ component field with the basis function of level $n$. So, the inner product between the rank--$n$ and rank--$(n-2)$ field in the term 
\begin{equation}
    \iint (\cosh z)^{2(n-1)} \phi'^{\mu(n-2)}L_{d+2n-4}L_{d+2n-6}\phi_{\mu(n-2)}\;,
\end{equation}
can be calculated just by using the orthonormal property of the basis functions \eqref{e:inn_prod_PP}. 

This is crucial in understanding the ansatz \eqref{e:ansatz_nn}--\eqref{e:ansatz_nn-2}. We assume that all inner products within the action must be evaluated just by means of the inner product formula \eqref{e:inn_prod_PP}. This assumption was proved to be correct since there exist such coefficients \eqref{e:A_n}--\eqref{e:I_n} that the ansatz \eqref{e:ansatz_nn}--\eqref{e:ansatz_nn-2} is gauge invariant under the gauge transformation \eqref{e:gauge_L}. Thus, each field must enter with the appropriate number\footnote{Not greater than two because the original \1ads \Fa contains two derivatives at most.} of raising and/or lowering $L$--operators in such a way that, eventually, after applying all $L$--operators the action ends up with basis functions of the same level in each term.   

This is the reason why there is no such terms as  
\begin{equation}
    S_{nn-4} = \iint (\cosh z)^{2(n-1)} \nabla_\mu\nabla_\nu\phi'^{\mu\nu\rho(n-4)}\phi_{\rho(n-4)}\;,
\end{equation}
describing interaction between the spin--$n$ and spin--$(n-4)$ fields. Even though this term could be written out of pure tensorial considerations it does not meet the requirement of eliminating the slicing coordinate solely via the inner product of the same level basis functions. 


\section{List of coefficients}\label{ap:An-In}

\begin{al}
  A_n =& \frac{s!(d+2s-4)!\Gamma(\frac{d}{2}+n-1)}{(s-n)!n!(d+n+s-4)!2^{s-n}\Gamma(\frac{d}{2}+s-1)}\;,\label{e:A_n}\\
  B_n =& -A_n\Bigg[ \frac{2(d+n+s-2)(d+2n-3)(s-n-1)}{d+2n-2} + n(d+n-2) \Bigg]\;,\label{Bn}\\
  C_n =& \frac{1}{2}n(n-1)A_n\Bigg[ (n-1)(d+n-3)-2 + \frac{(d+n+s-3)(d+2n)(s-n)}{2(d+2n-2)} \Bigg]\;,\\
  D_n =& -A_n\frac{(d+n+s-2)(s-n-1)}{d+2n-2}\;,\label{Dn}\\
  E_n =&  \frac{n(n-1)}{4(d+2n-4)}A_n\Bigg[\frac{(d+n+s-3)(d+2n)(s-n)}{d+2n-2} - 2(s-n+1)(d+n+s-4)\Bigg]\;,\label{En}\\
  F_n =& \frac{2(d+n+s-4)n}{d+2n-4} A_n\;,\\
  G_n =& -\frac{2(d+n+s-4)n(n-1)}{d+2n-4} A_n\;,\\
  H_n =& \frac{(d+n+s-4)n(n-1)(n-2)}{2(d+2n-4)} A_n\;,\\
  I_n =& \frac{(d+n+s-4)(d+n+s-5)n(n-1)}{(d+2n-4)(d+2n-6)} A_n\;.\label{e:I_n}
\end{al}

\begin{al}
  U_1 =&  \frac{s!(d+1)!(d+2s-4)!}{2(s-3)!(d+s-1)!(d+s-2)!}\;,\label{U}\\
    V_1 =& -\frac{(d+2s-4)!\Gamma(\frac{d}{2})}{(d+s-3)!2^{s-2}\Gamma(\frac{d}{2}+s-1)}s(s-2) \;,\\
  V_2 =& \frac{(d+2s-4)!\Gamma(\frac{d}{2})}{(d+s-3)!2^{s-2}\Gamma(\frac{d}{2}+s-1)}s(d+s-3) \;.\\
  W_1 =& \frac{(d+2s-4)!\Gamma(\frac{d}{2}+1)}{(d+s-2)!2^{s-2}\Gamma(\frac{d}{2}+s-1)}s(s-1)\;,\\
  X_1 =& - W_1\frac{s(s-1)(d+s-3)}{2d}(s-2)\;,\\
  X_2 =& - W_1\frac{s(s-1)(d+s-3)}{2d}((s-4)d-2s+4)\;,\\
  X_3 =& W_1\frac{s(s-1)(d+s-3)}{2d}\big( (d-2)^2+d(d+s-2)-2s \big)\;,\\
  X_4 =& W_1\frac{s(s-1)(d+s-3)}{2d}\big(-(d-2)^2 + d(d+s-2)+2s\big)\;,\\
  Y_1 =& \frac{(d+2s-4)!\Gamma(\frac{d}{2})}{(d+s-2)!2^{s}\Gamma(\frac{d}{2}+s-1)}(s-2)(ds+d+s-3) \;,\\
  Z_1 =& -\frac{(d+2s-4)!\Gamma(\frac{d}{2})}{(d+s-2)!2^{s}\Gamma(\frac{d}{2}+s-1)}\frac{1}{2}(s-1)(d+s-3) (s-2)^2 \;,\\
  Z_2 =& \frac{(d+2s-4)!\Gamma(\frac{d}{2})}{(d+s-2)!2^{s}\Gamma(\frac{d}{2}+s-1)}\frac{1}{4}(s-1)(d+s-3) d\big(s^2-4+d(s^2+4)\big) \;,\\ 
  Z_3 =& \frac{(d+2s-4)!\Gamma(\frac{d}{2})}{(d+s-2)!2^{s}\Gamma(\frac{d}{2}+s-1)}\frac{1}{4}(s-1)(d+s-3) (s-2)\big(8(s-2)+3d(s+2)  \big) \;.\label{Z}
\end{al}
\begin{al}
  &W = W_1 \sqrt{\frac{d}{d+1}}\;,\label{e:WXYZ}\\
  &X = \frac{(d+s-3)(ds+s-2)}{d}W\;,\\
  &Y = \frac{(d+2s-4)!\Gamma(\frac{d}{2}+1)}{(d+s-2)!2^{s-1}\Gamma(\frac{d}{2}+s-1)}\frac{(s-2)(ds+d+s-3)}{d+1}\;,\\
  &Z = \frac{(d+2s-4)!\Gamma(\frac{d}{2})}{(d+s-2)!2^s\Gamma(\frac{d}{2}+s-1)}\frac{(s-1)(d+s-3)}{d+1}\big( d^2(s-2)^2+d(3s^2-8s+4)+2(s-2)^2  \big)\;.\label{e:WXYZend}
\end{al}

\section{Higher--spin sector for the spin $s=4$ case}\label{s:s4}

In order to illustrate the approach described in Section \bref{s:hs}, in this Appendix we discuss  in detail the higher--spin sector of the EOM in the case of $s=4$ paying additional attention to the gauge parameters and the consistency condition. The higher--spin sector consists of independent subsectors of the following fields: $\phi^{\mu(2)}_{k}$, $\phi^{\mu(3)}_{k-1}$, $\phi^{\mu(4)}_{k-2}$ (for the fixed mode number $k\geqslant 2$)\footnote{The case $k = 1$ will be not considered here since it is trivial and can be analyzed analogously. }. Each subsector is invariant under the gauge transformations with the following gauge parameters: $\xi^{\mu(2)}_{k-1}$, $\xi^{\mu(3)}_{k-2}$. To simplify calculations and to make distinguishing between the fields and gauge parameters of different ranks easier, we introduce a new notation 
\begin{equation}
    h^{\mu(2)} = \phi^{\mu(2)}_{k}\;,\quad w^{\mu(3)} = \phi^{\mu(3)}_{k-1}\;,\quad q^{\mu(4)} = \phi^{\mu(4)}_{k-2}\;, \quad  \xi^{\mu(2)} = \xi^{\mu(2)}_{k-1}\;, \quad  \eta^{\mu(3)} = \xi^{\mu(3)}_{k-2}\;.
\end{equation} 

Using this notation the EOM of the higher--spin sector in $d=2$ read\footnote{Note that here each equation for the spin--$n$ field is divided by the constant $2A_n$ so that the coefficient of the first kinetic term $\Box \phi^{\mu(n)}$ is unit.}
\begin{al}
&\Box h^{\mu(2)} - \nabla^\mu\nabla_\nu h^{\mu\nu} + \nabla^\mu\nabla^\mu h + g^{\mu(2)}\nabla_\nu\nabla_\rho h^{\nu\rho}  - g^{\mu(2)}\Box h + \Big(\frac{3}{2}k(k+3) - 10\Big) h^{\mu(2)} + \notag\\
& + \Big(\frac{9}{4}k(k+3) + 11\Big) g^{\mu(2)}h + \sqrt{(k-1)k(k+3)(k+4)} q^{\mu(2)} -\sqrt{k(k+3)} \Big[ 2\nabla_\nu w^{\mu(2)\nu} - \notag\\
&- 2\nabla^\mu w^{\mu}  +g^{\mu(2)}\nabla_\nu w^{\nu}  \Big] = 0\;,\label{hsa}\\
&\Box w^{\mu(3)} - \nabla^\mu\nabla_\nu w^{\mu(2)\nu} + \frac{1}{2}\nabla^\mu\nabla^\mu w^{\mu} + g^{\mu(2)}\nabla_\nu\nabla_\rho w^{\mu\nu\rho}  - g^{\mu(2)}\Box w^{\mu} - \frac{1}{2}g^{\mu(2)}\nabla^\mu\nabla_\nu w^{\nu} - \notag\\
& - 9 w^{\mu(3)} + \Big(\frac{3}{2}k(k+3)+6\Big) g^{\mu(2)}w^{\mu} - \sqrt{(k-1)(k+4)} \Big[ \nabla_\nu q^{\mu(3)\nu} - \nabla^\mu q^{\mu(2)}  + \frac{1}{2}g^{\mu(2)}\nabla_\nu q^{\mu\nu}  \Big] +\notag\\
& + \frac{5}{4}\sqrt{k(k+3)} \Big[ \nabla^\mu h^{\mu(2)} - 2g^{\mu(2)}\nabla_\nu h^{\mu\nu} + \frac{1}{2}g^{\mu(2)}\nabla^\mu h \Big]= 0\;,\label{hsb}\\
&\Box q^{\mu(4)} - \nabla^\mu\nabla_\nu q^{\mu(3)\nu} + \frac{1}{2}\nabla^\mu\nabla^\mu q^{\mu(2)} + g^{\mu(2)}\nabla_\nu\nabla_\rho q^{\mu(2)\nu\rho}  - g^{\mu(2)}\Box q^{\mu(2)} - \frac{1}{2}g^{\mu(2)}\nabla^\mu\nabla_\nu q^{\mu\nu} - \notag\\
&-\big(k(k+3)-2\big) q^{\mu(4)} + \big(k(k+3) + 3\big)g^{\mu(2)}q^{\mu(2)} + \sqrt{(k-1)(k+4)}\Big[ \nabla^\mu w^{\mu(3)} - 2g^{\mu(2)}\nabla_\nu w^{\mu(2)\nu} +\notag\\
& + \frac{1}{2}g^{\mu(2)}\nabla^\mu w^\mu  \Big] + \frac{5}{4}\sqrt{(k-1)k(k+3)(k+4)}g^{\mu(2)}h^{\mu(2)} +\frac{1}{2}\sqrt{(k-1)(k+4)} g^{\mu(2)}g^{\mu(2)}\nabla_\nu w^\nu -\notag\\
&-\frac{5}{4}\sqrt{(k-1)k(k+3)(k+4)}g^{\mu(2)}g^{\mu(2)}h = 0\;,\label{hsc}
\end{al}
where the traces are denoted as $h = h^{\mu\nu}g_{\mu\nu}$, $w^{\mu} = w^{\mu\nu\rho}g_{\nu\rho}$ and $q^{\mu(2)} = q^{\mu(2)\nu\rho}g_{\nu\rho}$. Note that the last two terms in \eqref{hsc} arise from the double--tracelessness of the spin--4 field $q^{\mu(4)}$. The system of equations \eqref{hsa}--\eqref{hsc} is invariant under the following gauge transformations
\begin{all}\label{s4gauge}
&\delta h^{\mu(2)} = 2\sqrt{k(k+3)}\xi^{\mu(2)} \;,\\
&\delta w^{\mu(3)} = \nabla^{\mu}\xi^{\mu(2)} +  \sqrt{(k-1)(k+4)}\eta^{\mu(3)} \;,\\
&\delta q^{\mu(4)} = \nabla^{\mu}\eta^{\mu(3)}  - \frac{1}{2}\sqrt{(k-1)(k+4)} g^{\mu(2)}\xi^{\mu(2)} \;.
\end{all}

\begin{enumerate}
    \item Using the gauge transformations \eqref{s4gauge}, we impose the tracelessness gauge on the spin--3 and spin--4 fields $w^\mu = 0$, $q^{\mu(2)} = 0$. Such a gauge fixing drastically simplifies the EOM \eqref{hsa}--\eqref{hsc} that are gauge invariant under \eqref{s4gauge} with the gauge parameters satisfying the following conditions
    \begin{all}\label{cond}
    &\nabla_\nu\xi^{\mu\nu} = 0\;,\\
    &\nabla_\nu\eta^{\mu(2)\nu} = \frac{3}{2}\sqrt{(k-1)(k+4)}\xi^{\mu(2)}\;.
    \end{all}
    
    \item The Schouten identity \eqref{Schouten} along with the tracelessness condition allows us to express the spin--2 field $h^{\mu(2)}$ in terms of the divergence of the spin--3 field $\nabla_\nu w^{\mu(2)\nu}$ from the first EOM \eqref{hsa}
    \begin{equation} \label{h_nabla_w}
        h^{\mu(2)} = \frac{4\sqrt{k(k+3)}}{3\big(k(k+3)-8\big)}\nabla_\nu w^{\mu(2)\nu}\;,
    \end{equation}
    where the denominator does not vanish for all $k \geqslant 2$.
    
    \item Inserting the expression \eqref{h_nabla_w} into the trace of the second EOM \eqref{hsb} provides us with the equation on the double divergence of the spin--3 field 
    \begin{equation}\label{doubletr}
        -8(k+1)(k+2)\nabla_\nu\nabla_\rho w^{\mu\nu\rho} = 0\;,
    \end{equation}
    where the prefactor $-8(k+1)(k+2)\neq 0$ for any given $k\geqslant 2$. Thus, the spin--3 field is double--transversal on--shell.
    
    Now, we can additionally impose the transversal gauge on the spin--3 field. The equation on the gauge parameters reads 
    \begin{equation}\label{xi_nabla_w}
        \Box\xi^{\mu(2)} + \nabla_\nu\nabla^{\mu}\xi^{\mu\nu} +  \sqrt{(k-1)(k+4)}\nabla_\nu \eta^{\mu(2)\nu} = - \nabla_\nu w^{\mu(2)\nu}\;.
    \end{equation}
    This equation is consistent with the previous conditions \eqref{cond} arisen from the traceless gauge. It can be proved e.g. by taking the divergence of the both sides of the equation \eqref{xi_nabla_w}: the left--hand side vanishes because of the conditions \eqref{cond} and the right--hand side vanishes because of the double--tracelessness condition \eqref{doubletr}.
    
    Note that now the left--over gauge parameters satisfy the additional condition
    \begin{equation}\label{cond2}
        \Box\xi^{\mu(2)} + \nabla_\nu\nabla^{\mu}\xi^{\mu\nu} +  \sqrt{(k-1)(k+4)}\nabla_\nu \eta^{\mu(2)\nu} = 0\;.
    \end{equation}
    Using the previous conditions \eqref{cond} and the Schoten identity, one can find the left--over gauge parameter $\xi^{\mu(2)}$ from \eqref{cond2}
    \begin{equation}
        \frac{3}{2} \big(k (k+3)-8\big) \xi^{\mu(2)} = 0\;,
    \end{equation}
    where the prefactor does not equal zero for all $k \geqslant 2$. Thus, the rank--2 gauge parameter $\xi^{\mu(2)}$ is fully exhausted and can be excluded from further consideration.
    
    \item The transversality of the spin--3 field means that the right--hand side of the equation \eqref{h_nabla_w} equals zero and therefore the spin--2 field vanishes $h^{\mu(2)} = 0$. Applying the Schouten identity to the second equation \eqref{hsb} we can express the spin--3 field in terms of the divergence of the spin--4 field 
    \begin{equation}\label{w_nabla_q}
        w^{\mu(3)} = -\frac{1}{12}\sqrt{(k-1)(k+4)}\nabla_\nu q^{\mu(3)\nu}\;.
    \end{equation}
    Since the spin--3 field is transversal, the spin--4 field is double--transversal $\nabla_\nu \nabla_\rho q^{\mu(3)\nu\rho} = 0$. The double--transversality means that we can consistently impose the transversal gauge on the spin--4 field. Analogously to the previous step we check the consistency condition of this gauge fixing and find out that the rank--3 gauge parameter is exhausted as well $\eta^{\mu(3)} = 0$.  
    
    \item Since the spin--4 field is transversal, the spin--3 field vanishes $w^{\mu(3)} = 0$ from \eqref{w_nabla_q}. Finally, we can solve the last EOM \eqref{hsc}
    \begin{equation}
        -(k+1)(k+2)q^{\mu(4)} = 0\;,
    \end{equation}
    where the prefactor does not equal zero for all $k\geqslant 2$. Thus, the spin--4 field also vanishes $q^{\mu(4)} = 0$.

\end{enumerate}

\bibliographystyle{jhep}
\providecommand{\href}[2]{#2}\begingroup\raggedright\endgroup

\end{document}